\documentclass[12pt]{article}
\usepackage{latexsym}

\newcommand{\gam}{\gamma}

\newcommand{\ddg}{\ddagger}
\newcommand{\tl}{\tilde}

\newcommand{\oo}{\over}

\newcommand{\be}{\begin{equation}}
\newcommand{\bear}{\begin{eqnarray}}
\newcommand{\ear}{\end{eqnarray}}
\newcommand{\ee}{\end{equation}}
\newcommand{\lbl}{\label}
\newcommand{\bi}{\bibitem}
\newcommand{\ci}{\cite}

\newcommand{\vs}{\vspace}

\begin{document}

\

\baselineskip .7cm

\vs{27mm}

\begin{center}

{\LARGE \bf On an Alternative Approach to the Relation between Bosons and Fermions:
Employing Clifford Space}

\vs{3mm}

Matej Pav\v si\v c

Jo\v zef Stefan Institute, Jamova 39,
1000 Ljubljana, Slovenia

e-mail: matej.pavsic@ijs.si

\vs{6mm}

{\bf Abstract}

\end{center}

\vs{2mm}

We further explore the idea that physics takes place in
Clifford space which should be considered as a generalization of
spacetime.
Following the old observation that spinors can be represented as
members of left ideals of Clifford algebra, we point out that
the transformations which mix bosons and fermions could be
represented by means of operators acting on Clifford algebra-valued
(polyvector) fields. A generic polyvector field can be expanded
either in terms of bosonic or in terms of fermionic fields.
In particular, a scalar field can transform into a mixture of bosonic
and/or fermionic fields.

\vs{8mm}

The idea that Clifford algebra might provide a clue to the unification of
fundamental interactions has been explored by a number of researchers
\ci{CliffUnificOthers, Pezzaglia, Castro}, \ci{PavsicCliff}--\ci{PavsicArena}.
Crawford considered
local transformations which involve all $2^n$ generators of Clifford
algebra acting on column spinors. The compensating gauge fields behaved
as Yang-Mills fields. Instead of column spinors one can represent spinors
geometrically, as memebers of left or right ideals of Clifford algebra
\ci{Riesz, Teitler}. Chisholm  used geometric spinors, but he took only
spinor of one left ideal, and left out those belonging to other
left ideals. The first who took advantage of the full Clifford algebra
in describing wave functions was Pezzaglia. He proposed that wave functions
be Clifford algebra valued objects, called {\it polyvectors}, satisfying
the generalized Klein-Gordon or the
Dirac  equation. In his particular model the wave function depended on
vector and bivector coordinates. In ref.\,\ci{PavsicSchladming} the polyvector
Dirac equation in which the polyvector wave function depended on
polyvector coordinates, wheareas in ref.\,\ci{CastroFound} the analogous
wave function in the context of the Klein-Gordon equation was
considered. Those ideas were subsequently further developed in
refs.\,\ci{PavsicCliff, PavsicBook, CastroPavsicHigher}, and it was
realized that a polyvector field can contain either bosons or
fermions, or both at once, and thus provided a realization of a kind
of supersymmetry.

In this letter I will demonstrate how the transformations, proposed in
by Chisholm and Crawford \ci{CliffUnificOthers},
when operating on polyvector fields (``wave functions"),
can change bosons into fermions, and vice versa.

Let $X$ be a coordinate polyvector
\be
     X= x^M \gam_M = \sum_{r=1}^n x^{\mu_1 ...\mu_n} 
     \gam_{\mu_1 ... \mu_r} \; , \quad \mu_1 < \mu_2 < ... < \mu_r
\lbl{1}
\ee
expanded in terms of the basis $\lbrace \gam_M \rbrace = ({\bf 1}, \gam_{\mu_1},
\gam_{\mu_1 \mu_2},...,\gam_{\mu_1 ...\mu_r})$ of the Clifford algebra
generated by $\gam_\mu$, $\mu=1,2,...,n$ that satisfy
\be
      \gam_\mu \cdot \gam_\nu \equiv {1\oo 2} (\gam_\mu \gam_\nu + 
      \gam_\nu \gam_\mu)
      = g_{\mu \nu} {\bf 1}
\lbl{2}
\ee
The basis $\lbrace \gam_M \rbrace$ spans a manifold, called {\it Clifford
space}, shortly $C$-space.
The metric of $C$ space is defined as the scalar product
\be
      G_{MN} = \gam_M^\ddg * \gam_N
\lbl{3a}
\ee
Here `$\ddg$' denotes the reversion, that is the operation which reverses
the order of the generators $\gam_\mu$ (for example, 
$\gam_{\mu_1 \mu_2 \mu_3}^\ddg = \gam_{\mu_3 \mu_2 \mu_1}$), whilst `*' 
denotes the scalar product between
two Clifford numbers $A$ and $B$
\be
     A * B = \langle AB \rangle_0
\lbl{3b}
\ee
In general, C-space is curved and we distinguish between the coordinate
basis elements $\gam_M$ and local basis elements $\gam_A$ \ci{PavsicKaluzaCliff}.
For simplicity, in this paper we confine our consideration to {\it flat}
$C$-space, therefore $\gam_M=\gam_A$.

Let us consider a field theory in $C$-space, by introducing a polyvector
field
\be
    \Phi = \phi^M (X) \gam_M = \phi^A (X) \gam_A
\lbl{3}
\ee
We assume that the components $\phi^A$ are in general complex valued.
We interprete the imaginary unit $i$ in the way that is usual in quantum
theory, namely that $i$ lies outside the Clifford algebra of spacetime and
hence commutes with all $\gam_M$. This is different from the point of
view hold by many researchers of the geometric calculus based on Clifford
algebra (see, e.g., \ci{Hestenes, Lounesto}). They insist that $i$ has to
be defined geometrically, so it must be one of the elements of the set
$\lbrace \gam_A \rbrace$, such that its square equals $-1$. An alternative
interpretation, also often assumed, is that $i$ is the pseudoscalar unit of a 
higher dimensional space. For instance, if our spacetime
is assumed to be 4-dimensional, then $i$ is the pseudoscalar unit of a
5-dimensional space. The problem then arises about a physical intepretation
of the extra dimension. This is not the case that we adopt. Instead we
adopt the view, first proposed in \ci{PavsicBook}, that $i$ is the bivector
of the 2-dimensional {\it phase space} $P_2$, spanned by $e_q,~e_p$, so that
$Q\in P_2$ is equal to $Q=e_q e_q + p e_p,~ e_q Q = q + i p, ~ i=e_q e_p$.
So our $i$ is also defined geometrically, but the space we employ differs form
the spaces usually considered in defining $i$. Taking into account that
there are four spacetime dimensions, the total phase space is thus the
direct product $M_4 \times P_2 = P_8$, so that any element $Q \in P_8$ is
equal to $Q= x^\mu e_\mu e_p + p^\mu e_\mu e_p,~~e_q Q = (x^\mu + i p^\mu)
e_\mu$. This can then be generalized to Clifford space by replacing
$x^\mu,p^\mu$ by the corresponding Clifford space variables $x^M, p^M$.
In a classical theory, we can just consider $x^\mu$ only (or $x^M$ only), and
forget about $p^\mu$ ($p^M$), since $x^\mu$ and $p_\mu$ are independent. In
quantum theory, $x^\mu$ and $p_\mu$ ($x^M$ and $p_M$) are complementary
variables, therefore we cannot formulate a theory without at least implicitly
involving the presence of momenta $p_\mu$ ($p_M$). Consequently, wave
functions are in generally complex valued. Hence the occurrence of $i$ in
quantum mechanics is not perplexing, it arises from phase space.
We adopt here the conventional interpretation of quantum mechanics;
no hidden variables, B\" ohmian potential, etc., just the Born statistical
interpretation and Bohr-Von Neumann projection postulate.

Following Teitler \ci{Teitler} (whose work built on Riesz \ci{Riesz})
we introduce spinors as follows. Instead of the basis $\lbrace  \gam_A \rbrace$ one can consider another
basis, which is obtained after multiplying $\gam_A$ by 4 independent
primitive idempotents \ci{Teitler}
\be
    P_i = {1\oo 4} ({\bf 1} + a_i \gam_A + b_i \gam_B + c_i \gam_C) \; , \quad
   i=1,2,3,4
\lbl{4}
\ee
such that
\be
    P_i = {1\oo 4} ({\bf 1} +a_i \gam_A)({\bf 1}+b_i \gam_B) \; , 
    \quad \gam_A \gam_B
   = \gam_C \; , \quad c_i = a_i b_i
\lbl{5}
\ee
Here $a_i,~b_i,~c_i$ are complex numbers chosen
so that $P_i^2 = P_i$. For explicit and systematic
construction see \ci{Teitler, Mankoc}.

By means of $P_i$ we can form minimal ideals of Clifford algebra. A
basis of a left (right) minimal ideal is obtained by taking one of
$P_i$ and multiply it from the left (right) with all 16 elements
$\gam_A$ of the algebra:
\be
   \gam_A P_i \in {\cal I}_i^L \; , \qquad P_i \gam_A \in {\cal I}_i^R
\lbl{6}
\ee
Here ${\cal I}_i^L$ and ${\cal I}_i^R$, $i=1,2,3,4$ are four independent
minimal left and right ideals, respectively. For a fixed $i$ there are
16 elements $P_i \gam_A$, but only 4 amongst
them are different, the remaining elements are just
repetition of those 4 different elements.

Let us denote those different elements $\xi_{\alpha i}$, $\alpha=
1,2,3,4$. They form a basis of the $i$-th left ideal. Every Clifford
number can be expanded either in terms of $\gam_A =
({\bf 1},\gam_{a_1}, \gam_{a_1 a_2}, \gam_{a_1 a_2 a_3},
\gam_{a_1 a_2 a_3 a_4})$ or in terms of $\xi_{\alpha i} =
(\xi_{\alpha 1},~\xi_{\alpha 2},~\xi_{\alpha 3},~\xi_{\alpha 4})$:
\be
   \Phi = \phi^A \gam_A = \Psi = \psi^{\alpha i} \xi_{\alpha i} =
\psi^{\tilde A} \xi_{\tilde A}
\lbl{7}
\ee
In the last step we introduced a single spinor index ${\tilde A}$
which runs over all 16 basis elements that span 4 independent
left minimal ideals. Explicitly, eq. (\ref{7}) reads
\be
   \Psi = \psi^{\tilde A} \xi_{\tilde A} = \psi^{\alpha 1} \xi_{\alpha 1}
  + \psi^{\alpha 2} \xi_{\alpha 2} + \psi^{\alpha 3} \xi_{\alpha 3} +
\psi^{\alpha 4} \xi_{\alpha 4}
\lbl{8}
\ee
Eq.(\ref{7}) or (\ref{8}) represents a direct sum of four independent
4-component spinors, each living in a different left ideal ${\cal I}_i^L$.
The polyvector field $\Phi$ can be expanded
either in term of $r$-vector basis elements $\gam_A$ (bosonic fields)
or in terms of the spinor basis elements $\xi_{\tl A}$ (fermionic fields).

A generic transformation in $C$-space which maps a polyvector
$\Psi$ into another polyvector $\Psi'$ is given by
\be
   \Psi' = R \Psi S
\lbl{9}
\ee
where
\be
     R = 
     {\rm e}^{\gam_A \alpha^A}
     \quad {\rm and}
     \quad S 
     = {\rm e}^{\gam_A \beta^A}
\lbl{10}
\ee
Here  
$\alpha^A$ and $\beta^A$ are parameters of the transformation. Requiring
that the transformations
(\ref{9}) should leave the quadratic form $\Psi^{\ddg} * \Psi$ invariant.
So we have $\psi'^\ddg * \Psi' = \langle \psi'^\ddg \Psi' \rangle_S =
\langle S^\ddg \Psi'^\ddg R^\ddg R \Psi S \rangle_S = 
\langle \Psi^\ddg \Psi \rangle_S = \Psi^\ddg * \Psi$, provided that
$R^\ddg R = 1$ and $S^\ddg S = 1$. Explicitly, the quadratic form 
reads $\Psi^\ddg * \Psi = \psi^{* \tl A} \psi^{\tl B} z_{{\tl A}{\tl B}}$,
where $z_{{\tl A}{\tl B}}= \xi_{\tl A}^\ddg * \xi_{\tl B}$ is the
spinor metric.

The transformations that are usually considered are those for which
$\beta^A = -\alpha^A$, i.e., $R=R^{-1}$, but here we allow for more general
transformations (\ref{9}). They mix bosonic and fermionic field.
This can be seen on the folowing example. Let $\phi {\bf 1}$ be a
scalar valued field. Operating on it by a transformation $R$ from the
left we obtain a new field which is a mixture of fields of different grades
$r$:
\be
        \Phi' = R \phi \, {\bf 1} = \phi\, {\rm e}^{\gam_A \alpha^A} {\bf 1}
            = \phi \, {\rm e}^{\xi_{\tl A} \alpha^{\tl A}} {\bf 1}
            = \phi'^A \gam_A = \psi'^{\tl A} \xi_{\tl A}
\lbl{11}
\ee
 From a scalar we thus obtain a polyvector. But a polyvector can
written, according to eq.(\ref{7}), as a mixture of fermionic fields.
Therefore, our transformation $R$ acting on $\phi$ has a role of
a supersymetric transformations. For other $r$-vector fields, i.e.,
the fields with definite grade $r$, we have:
\bear
   R \phi^a \gam_a &=& \phi^a {C_a}^B \gam_B = \phi''B \gam_B = 
   \psi''^{\tl B} \xi_{\tl B}   \lbl{11a} \\
   R \phi^{a_1 a_2} \gam_{a_1 a_2} &=& \phi^{a_1 a_2} 
   {C_{a_1 a_2}}^B \gam_B = \phi'''B \gam_B = 
   \psi'''^{\tl B} \xi_{\tl B}  \lbl{11b} \\  
   &\vdots&
\ear
The above equations say that an $r$-vector field, $r=0,1,2,3,4$, transforms
into a superpositon of $r$-vector fields, which in turn is a superposition
of spinor fields belonging to different left ideals.

In examples (\ref{11})--(\ref{11b}) a bosonic field is transformed into
a mixture of fermionic fields. The inverse transformations are also possible
and they read:
\bear
     R \psi^{\alpha 1} \xi_{\alpha 1} &=& 
     \psi^{\alpha 1} {K_{\alpha 1}}^{\tl B} \xi_{\tl B} = 
     \psi'^{\tl B} \xi_{\tl B} = \phi'^B \gam_B \lbl{12a} \\ 
    R \psi^{\alpha 2} \xi_{\alpha 2} &=& 
     \psi^{\alpha 2} {K_{\alpha 2}}^{\tl B} \xi_{\tl B} = 
     \psi''^{\tl B} \xi_{\tl B} = \phi''^B \gam_B \lbl{12b} \\
     &\vdots&
\ear
In eqs.\,(\ref{11})--(\ref{12b}) we have particular cases of a general
transformation (\ref{9}) which transforms a polyvector field $\Phi$ into another
polyvector field $\Phi'$, or equivalently, a generalized spinor field
$\Psi$ into another generalized spinor field $\Psi'$:

Such view on spinors and supersymmetries has potentially profound
implications for further development of field theory, and in particular,
of string theory. The procedure that we described above, can be applied to
generalized point particles and strings as well. 

We can envisage that physical objects are described
in terms of $x^M = (\sigma, x^\mu, x^{\mu \nu},...)$. The
first straightforward possibility is to introduce a single
parameter $\tau$ and consider a mapping
\be
    \tau \rightarrow x^M = X^M (\tau)
\lbl{13}
\ee
where $X^M (\tau)$ are 16 embedding functions that
describe a worldline in $C$-space. From the point of view of
$C$-space, $X^M (\tau)$ describe a wordlline of a ``point
particle": at every value of $\tau$ we have a {\it point} in
$C$-space. But from the perspective of the underlying
4-dimensional spacetime, $X^M (\tau)$ describe an extended
object, sampled by the center of mass coordinates $X^\mu (\tau)$
and the coordinates
$X^{\mu_1 \mu_2}(\tau),..., X^{\mu_1 \mu_2 \mu_3 \mu_4} (\tau)$.
They are a generalization of the center of mass coordinates in the sense
that they provide information about the object 2-vector, 3-vector, and
4-vector extension and orientation\footnote{A systematic and detailed
treatment is in ref. \ci{PavsicArena}.}.

Instead of one parameter $\tau$ we can introduce two parameters $\tau$ and
$\sigma$. Usual strings are described by the mapping $(\tau,\sigma) \rightarrow
x^\mu = X^\mu (\tau, \sigma)$, where the embedding functions
$X^\mu (\tau,\sigma)$ describe a 2-dimensional worldsheet swept
by a string. This can be generalized to $C$-space. So we obtain generalized
strings, considered in refs.\,\ci{PavsicParis, PavsicSaasFee}, described
by polyvector variables $X^M (\tau, \sigma) \gam_M$. According to the
procedure described above, the latter polyvector can be expanded as
as sum of bosonic or fermionic fields. Altogether there are 16 real
degrees of freedom incorporated in  bosonic fields $X^M (\tau, \sigma)$, or
equivalently in fermionic fields $\theta^{\tl A} (\tau,\sigma)$.
According to this approach we do not
need a higher dimensional target spacetime for a consistent formulation
of (quantized) string theory. Instead of a higher dimensional space
we have Clifford space.

\end{document}